\documentclass[aps,eqsecnum,amsmath,twocolumn,nofootinbib]{revtex4}
\usepackage{graphics,setspace,epsfig,color}
\newcommand{\be}{\begin{equation}}
\newcommand{\ee}{\end{equation}}

\begin{document}

\title[title]{A new class of quantum bound states: diprotons in extreme magnetic fields}
\author{Danielle Allor, Paulo Bedaque and Thomas D. Cohen}

\affiliation{Department of Physics \\
University of Maryland\\College Park, MD 20742}

\author{Charles T. Sebens}

\affiliation{Department of Physics \\
Massachusetts Institute of Technology\\Cambridge, MA 02139}


\begin{abstract}
This paper considers the possibility that two charged particles with
an attractive short-ranged potential between them which is not
strong enough to form bound states in free space, may bind in
uniform magnetic fields. It is shown that in the formal limit where
Coulomb repulsion is negligible ($q \rightarrow 0 \, \, \text{and}
\, \, B_0 \rightarrow \infty \, \, \text{with} \, \, q B_0 \, \,
\text{fixed,}$ where $q$ is the charge and $B_0$ the field strength)
there always exists a bound state for a system of two identical
charged particles in a constant magnetic field, provided that there
is a short-range uniformly attractive potential between them.
Moreover, it is shown that in this limit {\it any} potential with an
attractive s-wave scattering length will posses bound states
provided that the range of the potential is much smaller than the
characteristic magnetic length, $r_0 \equiv (\frac{q
B_0}{4})^{-1/2}$.  For this case, the binding is computed
numerically.  We estimate the size of the magnetic field needed to
approximately reach a regime where the formal limit considered
becomes a good approximation to the dynamics. These numerical
estimates indicate that two protons in an extremely strong magnetic
field such as might be found in a magnetar will bind to form a
diproton.
\end{abstract}

\maketitle

\section{Introduction}
In nonrelativistic quantum mechanics, uniformly attractive
potentials in one or two dimensions always allow the existence of
bound states. While this is not generally true in three
dimensions, the presence of a magnetic field can alter the
situation considerably:  attractive potentials between two charged
particles with insufficient strength to bind in free space can
bind in the presence of a strong external magnetic field. This
novel class of quantum bound states is the focus of the present
paper.

Consider the nonrelativistic quantum dynamics of two charged
particles in a uniform magnetic field.   It is useful to
specialize to the case where the charge and mass of the two
particles are identical; in this situation the center-of-mass
motion decouples from the relative motion. We begin the analysis
working in a formal limit:
\begin{equation}
q \rightarrow 0 \, \, \text{and} \, \, B_0 \rightarrow \infty \,
\, \text{with} \, \, q B_0 \, \, \text{fixed.} \label{fixed}
\end{equation}
This limit has the property that the interaction between each
particle and the external magnetic remains of fixed strength while
the Coulomb repulsion between the particles is suppressed.

As a first example we consider a finite-ranged central-force
potential between the two particles which is attractive at all
distances but is not strong enough to bind the particles in the
absence of the magnetic field. In such circumstances there is a
general theorem that the two particles will necessarily have a
bound state even if the potential is not strong enough to bind in
free space. However, the emergence of bound states in magnetic
fields from potentials not strong enough to bind in free space is
more general than this. One important case where such binding is
guaranteed to happen in the limit $q \rightarrow 0 \, \,
\text{and} \, \, B_0 \rightarrow \infty \, \, \text{with} \, \, q
B_0 \, \, \text{fixed}$ is when  there is a finite ranged
interaction with both attractive and repulsive regions but which
has net attraction at low energies in the sense that the
scattering length, $a$, is attractive and the characteristic
magnetic length scale $(\frac{q B_0}{4})^{-1/2}$ (where we work in
natural units natural units of $\hbar=c=1$) is much larger than
the range of the interaction $R$.

There are simple intuitive reasons why such binding might be
expected. Consider first classical scattering in the situation
described. Qualitatively, if the Coulomb repulsion is neglected,
two non-interacting particles with the same charge and mass moving
in a magnetic field will have the same angular frequency. Thus
after they scatter they will bend in the same manner and thereby
remain close to one another for a longer time than an analogous
scattering in free space.  This in turn allows the attractive but
finite-ranged interaction to remain significant for a longer
period leading to increased net attraction.  The quantum analog of
this is presumably also increased attraction compared to the free
space case and it is not surprising that this can bring about
binding.

It can also be understood qualitatively from a quantum mechanical
perspective.  Recall that a charged particle in a constant
magnetic field of strength $B_0$ moves at the cyclotron frequency
$\omega_c = \frac{q B_0}{m}$ with an energy spectrum described by
the well-known Landau levels.  This can be easily seen via an
appropriate choice of gauge (the symmetric gauge) in which case
the Hamiltonian of a single charged particle in a magnetic field
can be written as the sum of a two-dimensional harmonic oscillator
Hamiltonian and an angular momentum term in the direction of the
magnetic field.  In this gauge, the magnetic field (which without
loss of generality we take to be in the $\hat{z}$ direction) acts
to restrict the particles to confined harmonic motion in the $x-y$
plane, leaving free particle motion only along the direction of
the magnetic field. Under these circumstances, it is highly
plausible that a situation similar to the one dimensional case
might arise in which any purely attractive potential between the
two particles will generically bind them.

Given the generality of the result, one might think  that such bound
states should be ubiquitous. However, there is a fundamental
limitation:  the analysis was carried out in the regime $q
\rightarrow 0 \, \, \text{and} \, \, B_0 \rightarrow \infty \, \,
\text{with} \, \, q B_0 \, \, \text{fixed}$.  In practice, however,
one cannot find particles with arbitrarily small charges since the
charge is quantized in units of $e$. Thus to approach the regime of
relevance one needs to find situations in which the Coulomb
interaction in the lowest lying state is very small compared to the
characteristic energies in the problem arising from the particles
interacting with the magnetic field, ({\it i.e.} the cyclotron
frequency---$\omega_c = \frac{q B_0}{m}$) and is also much smaller
than the characteristic energy scale of the short-ranged attraction
between the particles. In practice, these conditions are hard to
obtain. One requires {\it very} large magnetic fields.

The requirement of extreme magnetic fields implies that the effect
in its simplest form will not be directly relevant for systems of
atomic or molecular ions.  Note that, although one can describe the
interactions between such ions quantum mechanically via an effective
potential, the potential depends on the internal structure of the
ions.  The internal structure will be {\it greatly} altered due
polarization effects in very strong fields. Thus one cannot describe
the interaction of the two ions in the presence of the magnetic
field as essentially the free space interaction.   Of course, there
may well be situations in which two atomic or molecular ions bind in
the presence of a magnetic field which do not bind in free space.
However, such an effect would be a many-body effect in which the
effective two-body potential between the atoms is altered due to the
changed internal structure.  This is {\it not} the effect discussed
here. As we will show, the magnetic fields needed to see the
phenomenon discussed here are {\it many} orders of magnitude larger
than the characteristic field strength needed to alter the
interparticle potential by altering the internal structure

Thus, the most natural place to look for the effect is where strong
interactions dominate---i.e. in nuclear physics. The most favorable
case we know is for bound states of the two protons. As we will
discuss below, such diprotons are expected to form  in magnetic
fields of $~10^{12} \, {\rm T}$. Such fields {\it are} found in
nature but static fields of this scale are only seen in
astrophysical contexts such as magnetars. It remains an open
question as to whether there are any astrophysical consequences of
these states which have direct implication for observation. Whether
this turns out to be the case or not, these states are interesting
quantum mechanical curiosities in their own right.

Before proceeding it is useful to define what we {\it mean} by a
bound state in this context.  In free space this is unambiguous: the
relative and center of mass motion decouple, and  bound states have
a discrete energy spectrum, negative energy and a normalizable wave
function for the relative motion; the probability of finding the
particles at some fixed distance drops off exponentially with
distance. In contrast, continuum states in free space have a
continuum spectrum, positive energy and non-normalizable states. In
a magnetic field things are dramatically different. In the first
place, relative and center-of-mass motion do not generally decouple.
Secondly, the distinction between discrete and continuum spectra
becomes ambiguous: even non-interacting single particles have a
spectrum with discrete Landau levels (with infinite degeneracy)
along with a continuum index. Given this situation, we begin as a
technical matter with a  {\it definition} of a ``bound state'' in
terms of the behavior of the drop off of the probability amplitude
with distance: {\it a bound state is an energy eigenstate in which
the probability amplitude for finding two particles a distance $r$
apart falls of at long distances faster than a power law in all
directions}. This definition is, of course, fully consistent with
zero-field case and is a useful generalization.

In this paper, we restrict our attention to situations in which the
charge and mass of the two particles are identical.  As will be
discussed below, provided that the dynamics are non-relativistic,
the center-of-mass and the relative motion cleanly separate in this
case.  The center-of-motion behaves as a single particle of mass
$M=2m$ and charge $Q=2 q$; it has a spectrum of (infinitely
degenerate) discrete Landau levels plus a continuum associated with
the motion in the $\hat{z}$ direction. Bound states in this
circumstance have two properties analogous to the free space case:
i) the wave functions for the relative motions are normalizable and
ii) the energy of the state is negative by a non-zero amount
relative to what it would be if the particles were noninteracting or
equivalently were very far apart.

The goal of this paper is to explore the phenomenon of binding due
to intense magnetic fields rather generally.  We will show that in
certain circumstances the phenomenon must occur.  However, for any
particular interaction whether or not binding occurs and, if it
does, the value of the binding energy requires a detailed knowledge
of the interaction and a numerical solution of the Schrodinger
equation. It is useful to try to understand how this effect occurs
rather generally without relying on too many details of the
interaction (which in many cases is not known with precision in any
event).  Accordingly, we focus here on some limiting cases through
which one can gain insight into the phenomenon without specifying
the details of the particle-particle interaction.  The key
limitations in this analysis are a restriction to situations in
which the Coulomb interaction may be neglected and where the
magnetic fields are small enough so that the characteristic magnetic
length scale is much longer than the range of the potential.  While
these limitations are severe, there is at least one interesting
physical system---a bound state of protons in magnetar strength
fields---for which the approximations appear to be quite good.

The plan of the paper is as follows:  We begin by setting up the
general problem and then prove that, in a constant magnetic field
and in the limit $q \rightarrow 0 \, \, \text{and} \, \, B_0
\rightarrow \infty \, \, \text{with} \, \, q B_0 \, \,
\text{fixed}$, two charged particles with the same charge and mass
 and with a uniformly attractive potential between them always
have a bound state.  We also prove that in this regime {\it any}
potential which is attractive in the sense of having an attractive
scattering length must have a bound state in a sufficiently weak
magnetic field and derive an analytic expression for the binding
energy for weak magnetic fields.  In connection to this weak field
limit we note that it is very difficult to simultaneously have
sufficiently weak fields to be in this regime and to have the
Coulomb repulsion be small enough to neglect.

Next we consider systems for which the scattering length is
``unnatural'', {\it ie.}  much larger then the characteristic
range of the interaction. In this regime there is an effective
field theory \cite{kaplan,burgess} describing the system and one can compute
binding without a detailed microscopic knowledge of the
interaction.  This regime is of phenomenological interest in low
energy nuclear physics since the nucleon-nucleon scattering length
is much larger than the characteristic range of the nuclear force.
We numerically calculate the binding energy of the bound state in
this effective theory as a function of a dimensionless parameter
proportional to $a B_0^{1/2}$ where $a$ is the scattering length
and $B_0$ is the strength of the magnetic field.  We discuss
corrections to this result both due to Coulomb repulsion and to
short-distance effects.  For the case of two protons we use
dimensional arguments to argue that a regime exists in which both
of these corrections are simultaneously moderately small.

Following this we  discuss the case where the two particles have
spin.  There is a general argument that energy of interaction of the
spins with the magnetic field are of the same order as the binding
effects considered here.  It is easy to see that in the regime of
validity of the effective theory the spin degrees of freedom
decouple from the spatial ones.  This means that the bound states
considered here will continue to exist.  However, in the case of
identical fermions it may well happen that even when there is a
bound state of the sort discussed in this paper, the ground state of
the system need not be bound. This can come about because the Pauli
principle combined with the dynamics of the bound state can require
the spins to be in distinct $m$-states. The energy gained in binding
can be less than the energy lost by aligning the spins in the field.
This situation is radically different from what occurs in free space
where the ground state must be bound if any bound states exist. As
we will see later, the situation where the lowest bound state is
{\it not} the ground state is precisely the circumstance for the
diproton.

We end the paper with a brief concluding section including a few
words about the possible astrophysical significance.

\section{Bound states in  magnetic Fields \label{BS}}

\subsection{Generalities}

In this section we set up the general problem and prove the
existence of a bound state when there is a potential which is
attractive everywhere between two identical charged particles in a
constant magnetic field in the limit $q \rightarrow 0 \, \,
\text{and} \, \, B_0 \rightarrow \infty \, \, \text{with} \, \, q
B_0 \, \, \text{fixed}$.

Consider the case of two charged nonrelativistic  particles (with
the same mass $m$ and the same charge $q$) interacting via a
central potential.  For simplicity we will neglect spin effects
here and return to the question of spin later in the paper;
accordingly we will assume a central potential initially. For
simplicity we will assume that the two particles have identical
charges and masses. The Hamiltonian of the system is then \be
    \hat{H} = \frac{(p_1 - q \hat{A})^2}{2 m}+\frac{(p_2 - q \hat{A})^2}{2 m} +
    V\left ( \left |\vec{r}_1 -\vec{r}_2 \right| \right) \, .
    \label{eqn_orig_hamiltonian}
\ee

Without loss of generality we can take the magnetic field to be in
the $\hat{z}$ direction. We find it convenient to work with the
symmetric  gauge
\begin{equation}
\vec{A} = \frac{B_0}{2}(x\hat{y}-y\hat{x}) \; .\label{gauge}
 \end{equation} For the case of interest here where the mass and
charge of the two particles are equal, the center-of-mass and
relative motion separate. With $\mu = \frac{m}{2}$ as the reduced
mass and $M = 2 m$ as the total mass, the relative and
center-of-mass Hamiltonians in configuration space are:
\begin{align}
    \hat{H}_{rel}=-\frac{1}{2 \mu} \nabla^2+\frac{1}{2} \mu \omega ^2 \left(x^2+y^2\right)-\omega
    L_z +V(|r|)
    \label{eqn_relhamiltonian}
    \\
    \hat{H}_{CoM}=-\frac{1}{2 M} \nabla^2+\frac{1}{2} M \omega ^2 \left(x^2+y^2\right)-\omega
    L_z,
    \label{eqn_CoMhamiltonian}
\end{align}
 where $\omega = \frac{1}{2} \omega_c = \frac{q B_0}{M} = \frac{q B_0}{4 \mu}$.

In the absence of a short-range potential, the center of mass and
the relative equations closely resemble the Hamiltonian of a single
particle with a Landau level spectrum.  In the absence of
interactions between the particles the relative motion does as well.
Therefore, without the potential, the total energy for the ground
state of the relative motion is $\omega = \frac{q B_0}{4 \mu}$. A
bound state corresponds to an energy below this value.

It is easy to convince one's self that to minimize the energy the
term proportional to $L_z$ in Eq.~(\ref{eqn_relhamiltonian})
should be zero. It is always possible to do this since $L_z$
commutes with the Hamiltonian. With the choice of gauge of
Eq.~(\ref{gauge}) the Hamiltonian can be written: \be
    \hat{H}_{rel}
    = -\frac{1}{2 \mu} \nabla^2+\frac{1}{2} \mu \omega ^2
    \rho ^2-V\left(\vec{r}\right),
    \label{eqn_relhamiltonian2}
\ee where  $V(\vec{r} $) is the
 potential, and $\nabla^2$ is the three
dimensional Laplacian.
A two-dimensional harmonic oscillator term, $\hat{H}_{2D-HO}$, may
be isolated from Eq.~(\ref{eqn_relhamiltonian2}) and the
Hamiltonian may be rewritten as follows: \be
    \hat{H}_{rel} = \hat{H}_{2D-HO}-\frac{1}{2 \mu} \frac{\partial^2}{\partial
    z^2}-V(\vec{r}).
    \label{eqn_2dho}
\ee

\subsection{A theorem}

A variational argument may then be used to show that a uniformly
attractive potential always binds with the Hamiltonian in
Eq.~(\ref{eqn_2dho}).  Before doing this note that the condition
in Eq.~(\ref{fixed}) is implicit in taking a uniformly attractive
interaction;  there will always be a long ranged repulsive Coulomb
potential away from this limit.  To proceed we take a simple a
wave function ans\"atz of  \be
    |\psi\rangle=|0\rangle_{2D-HO}
    |\phi\rangle_Z
    \label{eqn_inbasis}
    \; ,
\ee where $|0\rangle_{2D-HO}$ is the ground state of the
two-dimensional harmonic oscillator in the $x-y$ plane and $|\phi
\rangle_z$ is an arbitrary state in the Hilbert space for motion in
the $z$ direction with $\phi(z)=\langle z|\phi\rangle_z$; an outer
product between is implied. The expectation value of the energy in
this state is:
\begin{eqnarray}
  &{}&  \langle \psi | \hat{H}_{rel} | \psi \rangle   \nonumber
    \\
   & = & \omega \, +  \,\int d z \, \phi^*(z)( - \frac{1}{2 m} \frac{\partial ^2}{\partial z^2}
   -  V_{\rm eff}(z) ) \phi(z)  ,
    \label{eqn_energy} \\
&{}& {\rm with} \; \; V_{\rm eff}(z) \equiv \int \, dx \, dy
\psi^*_0(x,y)( V\left (\vec{r} \right ) ) \psi_0(x,y) \nonumber
\end{eqnarray}
where $\omega$ is the ground state energy of the two-dimensional
harmonic oscillator and $\psi_0(x,y)$ its normalized wave function.
Note that $\psi^*_0\psi_0$ is non-negative; thus, if $V(\vec{r})$ is
non-negative for all $\vec{r}$ then $V_{\rm eff}$ is non-negative
for all $z$.  That is, a purely attractive $V$ implies a purely
attractive $V_{\rm eff}$.

Note that the integral in $z$ {\it is} the expectation value for the
energy of a one-dimensional system with potential $V_{\rm eff}$. The
effective one-dimensional system must have at least one normalized
bound state with negative energy since there is a theorem that {\it
all} purely attractive one dimensional systems do \cite{1D}. Thus,
we can choose $\phi(z)$ to be the normalized ground state of this
system. In that case we have $\langle \psi | \hat{H}_{rel} | \psi
\rangle =\omega - B_G^{\rm eff}$ where $B_G^{\rm eff}$, the binding
energy of the effective one-dimensional problem, is positive.  Thus
we have found a variational state with energy less than that of two
non-interacting particles in the magnetic field; the true state must
be lower still and thus a bound state must exist.

\section{The weak field limit\label{WF}}
The theorem in the preceding section is of interest as it
illustrates the general phenomenon of particles which do not bind in
free space forming bound states in magnetic fields.  It is limited
in that it does not tell us whether a potential which has both
attractive and repulsive regions binds in a magnetic field.  As a
general rule one cannot determine whether this occurs for a
particular potential and a given field strength without solving the
Schr\"odinger equation explicitly.

However, there is one general situation where the existence of bound
states can be determined by general arguments: the case in which the
natural length scale associated with Landau orbits in the magnetic
field, $r_0 = \frac{1}{\sqrt{\mu \omega}}$, is much larger than the
range of the potential between the particles (which we will denote
as $R$).  In this weak field situation, it is easy to show that
bound states exist providing the scattering length is attractive and
Coulomb repulsion is negligible. Moreover, if the scattering length,
$a$ is ``natural'' ({\it i.e.} of order $R$) and thus much smaller
than $r_0$, one can directly compute the binding energy. It is given
by:
 \be
    E_b = 2 \left( \frac{a}{r_0} \right) ^2 \omega .
\label{Be} \ee

We note that the weak field condition $r_0 \gg R$ can be at odds with
the regime in which the formal limit  $q \rightarrow 0 \, \,
\text{and} \, \, B_0 \rightarrow \infty \, \, \text{with} \, \, q
B_0 \, \, \text{fixed}$ (in which Coulomb repulsion may be
neglected) can be regarded as approximately satisfied.  The
neglect of Coulomb requires $B_0$ to be sufficiently large; the
condition $r_0 \gg R$ requires $B_0$ to be small.  For the problem
of the diproton there does exist a window in $B_0$ for which the
system is in both regimes.

The weak field regime is of interest in that the system is
insensitive to the detail of the potential---at leading order, all
that is relevant is the scattering length.  This is precisely the
situation for which an effective field theory \cite{kaplan,burgess} is useful.
We begin by considering the ``natural scattering length'' regime
of $a \sim R$.  We consider the unnatural case $a \gg R$ in a
subsequent section.

Our goal is to demonstrate the relationship between the binding
energy and the scattering length of Eq.~(\ref{Be}).  We begin with
a general expansion of the bound state in a convenient basis: \be
    |\psi \rangle=\sum_{n}|n\rangle_{2D-HO}|\phi_n\rangle_z
    \label{eqn_inbasis2}
    \ee
where as before an outer product is implied;  $|\phi_n \rangle_z$
represents the particular $z$ space wave function coupled to a
given state of the two dimensional oscillator.  The key to our
demonstration here is that in the limit $a,R \ll r_0$ only the
$n=0$ term contributes.  We represent the total energy of the
system as $\omega$, the energy of the two dimensional harmonic
oscillator, minus the quantity of interest:
   $ \hat{H}_{rel}|\psi\rangle = \left(-E_b+\omega\right)|\psi\rangle
    $.
Using the expansion of Eq.~(\ref{eqn_inbasis2}) and multiplying
Eq.~(\ref{eqn_2dho}) on the left by the two dimensional harmonic
oscillator state $\langle m(\rho)|$ yields:
\begin{align}
    \sum_{n} \left( -\frac{1}{2 \mu} \frac{\partial^2}{\partial
    z^2} \delta_{mn} - V_{mn}(z) \right) \phi_n (z)  \notag
    \\
    = \sum_n
    \left( -E_b -n\omega \right)\phi_n (z)   \delta_{mn} ,
    \label{eqn_sumwithdeltas}
\end{align}
where $V_{mn}(z) \equiv \int   d\rho \, 2 \pi \rho \,
\psi^*_m(\rho)V(\rho,z) \psi_n(\rho)$.

Note that the solution to the Schr\"{o}dinger equation along the
$z$-axis decays exponentially according to $e^{-\sqrt{2 \mu E_b}
z}$ as the binding becomes small. Clearly, the $n=0$ state is the
only one which contributes in the weak binding limit because of
the $(-E_b -n \omega)$ term; for nonzero $n$, the wavefunction no
longer decreases exponentially in the $\hat{z}$ direction as the
binding energy becomes very weak. The only term remaining that
connects orthogonal $m$ and $n$ states is $V_{mn}$. However, the
range of the attractive potential $V$ is small compared to that of
the harmonic oscillator. Therefore, all harmonic oscillator states
appear essentially flat in the range of the potential, and all
contribute approximately equally.

Note that there are now three length scales in the problem: the
range of the attractive potential, $R$; the length scale of the
harmonic oscillator, $r_0 = \frac{1}{\sqrt{\mu \omega}}$; and the
scale of exponential decay of the wave function in the $\hat{z}$
direction, $\kappa^{-1} = (2 \mu E_b)^{-1/2}$. The hierarchy of
length scales needed for Eq.~(\ref{eqn_deltaschrodinger}) to be
valid is \be
    R \ll \, r_0 \ll \kappa^{-1} \label{cond1}.
\ee One must test {\it ex post facto} to see whether this
condition is satisfied.  Note that if Eq.~(\ref{Be}) is correct,
then the condition in Eq.~(\ref{cond1}) becomes

\be
R \ll r_0 \ll r_0 \frac{1}{a\sqrt{\mu\omega}}.
\ee

Eq.~(\ref{eqn_sumwithdeltas}) now takes the form of a
one-dimensional Schr\"{o}dinger equation.  Moreover, at small
binding the spatial extent of the one dimensional wave function
means that the system can be well represented as a system with a
Dirac delta potential in $z$ with strength $\alpha$ and energy
$-E_b$: \begin{eqnarray}
    \left( -\frac{1}{2 \mu} \frac{\partial^2}{\partial
    z^2} - \alpha \delta (z) \right)|\phi_0 (z) \rangle & = & -E_b |\phi_0 (z)
    \rangle, \nonumber \\
 {\rm with} \; E_b & = & \frac{\mu \alpha^2}{2} \; .
    \label{eqn_deltaschrodinger}
\end{eqnarray}   Clearly, to proceed we must determine $\alpha$ in terms of
the scattering length.

We know {\it a priori} that $\alpha$ does not depend on the
details of the short distance potential but only on the scattering
length.  This suggests a useful strategy for computing $\alpha$ in
general, namely to consider {\it any} class of potential for which
we know how to compute both $\alpha$ and the scattering length and
then relate the two.  There is a straightforward situation for
which we can do this: namely the case of arbitrarily weak
potentials.  For the bound state problem in a magnetic field this
regime gives
 \be \alpha=\int dz \, 2 \pi \, \rho \, d\rho \, V(\rho,z) \,
    \psi_0^*(0) \psi_0(0) \label{alpha} \ee
For the scattering, the weak potential limit justifies the the
Born approximation.  The Born approximation for the scattering
length is \be a = 4 \pi \mu \int d^3 x V \; .\label{born} \ee
Combining Eqs.~(\ref{eqn_deltaschrodinger}), (\ref{alpha}) and
(\ref{born}) yields Eq. (\ref{Be}) .

A couple of quick comments about this result are in order.  The
first is that Eqs.~ (\ref{alpha}) and (\ref{born}) only hold for
weak potentials. Never-the-less the result in Eq. (\ref{Be}) is
fully general for $R,a \ll r_0$. The reason is clear---we have an
effective theory in which the {\it only} relevant information
about the potential is through the scattering length.  How the
scattering length arises microscopically is irrelevant.  The
second is that this result generalizes the theorem considered
above.  {\it Any} potential with attractive scattering length will
form a bound state in a magnetic field provided that the system is
in a regime in which the Coulomb repulsion is negligible and the
range of the potential and its scattering length are much smaller
than $r_0$.

It is important to note however, that this weak coupling limit is
of more theoretical then practical significance.  Recall that the
analysis depended on two assumptions.  The first was that the
Coulomb repulsion was negligible and the second that the magnetic
length scale is much longer then all scales associated with the
potential including the scattering length.  Clearly neither
assumption is exact and hence there must be corrections to
these.   One can estimate the size of these corrections.

The characteristic distance between the particles is $(\mu
E_b)^{-1/2}$ since the system is weakly bound and the wave
function is largely beyond the range of the potential. From this
one estimates the characteristic scale of the perturbative Coulomb
correction to be
 \be
 \frac{\Delta E_b^{\rm Coulomb}}{E_b}
\sim {\cal O}\left(  \, \frac{q^2 \, \mu \, r_0^2 }{a}  \, \right )
\sim  {\cal O}\left ( q^2 (\mu R) \frac{r_0^2}{R^2} \right )
\label{deltaE_Cwc}\ee where $q$ is the charge and in the second form
we are taking the scattering length to be of the same order as the
range of the potential as one expects for a ``natural'' ({\it i.e.}
not fine tuned) problem. It is nontrivial to find a regime for which
this is small enough to be considered to be a small correction. In
the first place note that the second expression for the ratio of the
correction to the dominant part is proportional to $q^2 \mu R$. For
two light nuclei and strong interactions  $q^2 \mu R \sim
\frac{Z^2}{5 A}$. In comparison, in atomic systems $\alpha \mu R $
is many orders of magnitude larger. As noted earlier this implies
that binding of the sort considered in this will not occur in atomic
or molecular ions.  In fact, it is typically marginal at best for
nuclear systems in the weak coupling regime. Note that
Eq.~(\ref{deltaE_Cwc}) contains a factor of $r_0^2/R^2$.  Recall
that the weak field regime has $r_0^2/R^2 \gg 1$.  Thus, there is at
best a very small window, and more likely no window at all, for
which the system  has a natural scattering length and simultaneously
is at weak field and has perturbative Coulomb corrections.
Fortunately, there are systems with unnatural scattering lengths
where  binding due to the magnetic field can occur.

\section{Unnatural scattering lengths \label{unnat}}

In the previous section we considered binding in the regime where
both the range of the potential and the scattering length are much
smaller than the characteristic magnetic length scale.  In this
section, a more general case is considered: one in which as before
$r_0 \gg R$ so that the detailed structure  of the short-ranged
potential remains unimportant but where we no longer require the
scattering length to be much shorter than $r_0$.  Of course, one
generally expects that in the absence of fine tuning the scattering
length should be of the same scale as the range of the potential (or
less in the case of weak potentials). Under such circumstances $r_0
$ is also much larger than $a$. However, there are systems for which
the scattering length is ``unnaturally'' large, {\it i.e.} much
larger than the range of the potential. The case of unnatural
scattering lengths is critical in nuclear physics since the
nucleon-nucleon scattering lengths are unnatural:  $a \sim 20 \,
{\rm fm}$ while $R \sim 2 \, {\rm fm}$ \cite{deswart}. Indeed, much
of the development of effective field theory methods in nuclear
physics in the past decade has concerned the treatment of systems
which are unnatural \cite{bedaque,bira,all,daniel}. It is important
as a practical matter to consider the regime of unnatural scattering
lengths.  As noted above the regime considered previously with $r_0
\gg a$  turns out to be of very limited utility.

To proceed it is useful to reexpress the problem by adding and
subtracting a one dimensional harmonic potential in the $\hat{z}$
direction from the Hamiltonian in Eq.~(\ref{eqn_relhamiltonian}).
The Hamiltonian can be rewritten as a three dimensional harmonic
oscillator Hamiltonian minus a one dimensional harmonic
oscillator potential term in the $\hat{z}$ direction: \be
    \hat{H}_{rel}=-\frac{1}{2 \mu} \nabla^2+\frac{1}{2} \mu \omega^2 r^2-\frac{1}{2}
     \mu \omega^2 z^2 -V(|r|)
    \label{eqn_rewrittenhamiltonian} \; . \ee
This formulation is convenient for numerical studies in that one
can easily implement the calculation in a truncated three
dimensional harmonic oscillator basis.

As a first step towards considering the general case with $r_0 \gg
R$  but with the ratio of $r_0/a$ arbitrary, we consider the
extreme case in which the short range potential in the absence of
the magnetic field has a scattering length sufficiently long to be
treated as though it were infinite scattering  ({\it i.e.} as
though the system had a zero energy bound state corresponding to
the boundary between having bound and unbound states in the
absence of a magnetic field).  This regime is of interest
theoretically in that the potential introduces no scales into the
problem---its range is irrelevant since it is so short as to be
taken to be zero while the scattering length is irrelevant as it
is so long as to be effectively infinite.  In such a system the
binding energy {\it must be} proportional to $\omega$.   We wish
to find the constant of proportionality.

Clearly, the entire effect of the short-range potential on the
wave function for $r_0 \gg R$ can be encoded as a boundary
condition for the s-wave part of the wavefunction at small $r$.
One way to see this is to start with a finite ranged potential and
consider it in the limit in which the strength of the potential,
$V_0$, goes to infinity and the range goes to zero in such a
manner as to keep the scattering length fixed.  Since the
range of the potential is taken to zero (to reflect the underlying
short range of the potential compared to relevant scales) the
boundary condition is at $r=0$. Moreover the zero-range nature of
the interaction implies that the boundary condition is insensitive
to energy thus the zero energy behavior encoded in the scattering
length is sufficient. The condition is
 \begin{eqnarray}
     \frac{u_0'(r)}{u_0(r)} \Big\vert_{r=0} & = &\frac{1}{a}   \label{eqn_fscattering}\\
     \nonumber \\
     {\rm with} \; \; u_0(r)& =& \frac{r \, \int d \Omega \, \psi(r)}{4
 \pi} \nonumber
\end{eqnarray}

We consider the infinite scattering length case at the outset, so
this limiting procedure simply yields the logarithmic derivative of
the wave function with respect to $r$ approaching zero at $r=0$.
Diagonalizing the Hamiltonian for this system in a truncated basis
and testing the numerical convergence of the truncation yields a
binding energy of
 \be E_b = F_{\infty} \,
\omega  \; \; {\rm with} \; F_\infty \approx .60486 \label{asy}
\ee where the result is numerically accurate to the quoted
precision. As advertised the binding energy is simply proportional
to the spacing of the Landau levels of the free particles,
$\omega$.

\begin{figure}
   \centering
       \includegraphics[width=3.5in]{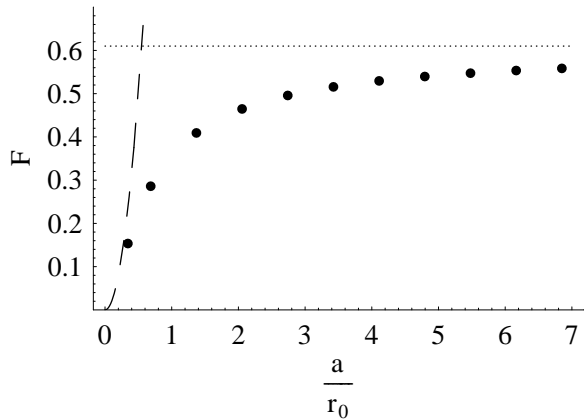}
       \caption{ The ratio of the binding energy to $\omega$ from Eq.~(\ref{f}), F,  as a function of $a/r_0$.
       The solid dots are the results of  numerical calculations.
       The dotted line indicates the asymptotic value from Eq.~(\ref{asy}) while the
       dashed line is the weak field result of Eq~(\ref{Be}).
       \label{fig_bindingenergy}}
\end{figure}

Of course, an infinite scattering length comes about only through
infinite fine tuning.  In practice, even when the scattering length
is quite large compared to the range of the interaction (as for
nucleon-nucleon scattering) it is finite.  This can play a
significant role since the external magnetic field may be varied
which in turn varies $r_0$.  Thus by changing the external field one
can go from the region in which $a \gg r_0$ to one in which $a \sim
r_0$.  It is important to consider the situation in which $a \sim
r_0 \gg R$   since this broadens the applicability of the analysis.
On general dimensional grounds it is apparent that if one uses
$\omega$ to set the scale of the binding energy, then then binding
energy can only depend on the dimensionless ratio $a/r_0$: \be E_b =
F(a/r_0) \omega \label{f} \; .\ee  We know from Eq.~(\ref{Be}) that
for small values $a/r_0$, $ F(a/r_0) \rightarrow 2 a^2/r_0^2$.
Similarly we know from Eq.~(\ref{asy}) that as $a/r_0$ gets large
$F$ approaches $ F_\infty \approx .60486$. The function $F$ can be
computed numerically by diagonalizing the magnetic Hamiltonian for
the subject to the boundary condition in
Eq.~(\ref{eqn_fscattering}). This is shown in
Fig.~\ref{fig_bindingenergy}.

The analysis in this section is more general than the one in
Sec. \ref{WF}.  We have shown that in circumstances where the
range of the potential is much shorter than the magnetic length
$r_0$ one always has a bound state if the scattering length is
attractive and if Coulomb repulsion can be neglected, and we have
explicitly computed the binding energy in this situation.

Clearly there are errors associated with approximating the
potential as zero-ranged and neglecting Coulomb repulsion.  If one
is in the unnatural regime where $a \sim r_0 \gg R$ one can
estimate the characteristic size of these errors.  From general
power counting arguments in effective field theory one expects
that relative errors induced by treating the potential as having
zero range are
 \be\frac{\Delta E_b^{\rm finite-range}}{E_b} \sim {\cal O}\left( {R^2}/{r_0^2} \right ) \; .
  \label{deltaE_fr} \ee
The relative errors due to the neglect of Coulomb are
  \be\frac{\Delta E_b^{\rm Coulomb}}{E_b} \sim {\cal O}\left(Q^2
   \, \mu r_0 \,
Log(R/r_0)  \right )  \label{deltaE_C} \ee .

There is a tension between the two approximations which is seen
clearly from Eqs.~(\ref{deltaE_fr}) and (\ref{deltaE_C}).  While
$\Delta E_b^{\rm finite-range}$ increases with increasing magnetic
field, $\Delta E_b^{\rm finite-range}$ decreases.  The optimal value
of the magnetic field to minimize both errors will have the two
types of errors as being of similar scales. Equating the two yields
an error at the optimal field strength for the two approximations of
$\Delta E_b/E_b \sim (\alpha \mu R)^{2/3}$. Plugging in values
typical for the range of the potential yields an error of order 10\%
for the diproton.  Thus one expects the analysis done here should
have a regime of approximate validity for the diproton case. While
it is possible that the corrections are large enough as to yield
numerically significant corrections to the result in
Fig~\ref{fig_bindingenergy}, it seems quite likely that a region
does indeed exist for which  two protons bind.

The preceding error estimate was done assuming the scattering length
is unnatural and of order $r_0$.  When the scattering length is
natural ({\it i.e.} $a \sim R$) the relative size of the Coulomb
correction is much larger and given by Eq.~(\ref{deltaE_Cwc}).
Clearly the two results must smoothly interpolate into one another.
It is easy to understand qualitatively how this comes about. If one
starts with the extreme limit of $a=\infty$ the error estimate in
Eq.~(\ref{deltaE_C}) remains valid since then $r_0$ sets the full
length scale in the problem.  As one decreases $a$ with fixed $r_0$
two effects occur: the binding energy drops as seen from
Fig.~\ref{fig_bindingenergy}.  This in turn leads to a larger wave
functions and hence smaller Coulomb corrections.  However, the
Coulomb energy drops much more slowly then the binding energy.

Ideally, one would like to calculate this in the regime of validity
of an effective field theory treatment---{\it i.e.} $r_0 \gg R$, $a
\gg R$ with $r_0/a$ arbitrary.  Unfortunately, this can not be done
without adding additional information about the system at short
distance.  If one took seriously our zero-range approximation for
the interaction, the perturbative estimate of the Coulomb
contribution would diverge  logarithmically at short distances due
to the singular nature of the Coulomb interaction. Of course, in
nature this behavior is regulated by the short distance physics of
the interaction.  This yields a logarithmic dependence of the ratio
of the short distance length associated with the potential and the
natural magnetic length scale. Thus, it is apparent that one needs
more data about the interaction than merely the scattering length to
compute the Coulomb energy.  This is unfortunate in that one would
like to test the qualitative estimates given in this section to
ensure that there are not anomalously large numerical coefficients
which might spoil the window where both approximations are
reasonably good.  We have done simple model calculations to explore
this possibility.  The models are consistent with our qualitative
estimates; none of the models raises doubts about whether two
protons bind in the regime where $r_0/a \sim .2$.

\section{Spin and the ground state \label{spin}}
Heretofore we have neglected spin in our discussion.  Of course, for
the general case particles with non-zero spin, the inter-particle
interactions will be spin dependent ({\it e.g.} there can be a
tensor force). As noted previously, for a generic interaction
whether or not a magnetic field will yield binding depends on the
details of the interaction and must be assessed on a case by case
basis through direct simulation.  As in the case of spinless
particles, if the system is in a regime in which the Coulomb
interaction can be neglected and the potential is attractive in all
channels and all distances a bound state must occur.  The previous
variational argument holds with only minor modification. The only
new issue is that in setting up a variational ans\"atz, the state
should be put into  some particular spin state consistent with the
statistics of the problem.

Similarly for $r_0 \gg R$, the analysis of the previous sections
should hold for the energy of a bound state both for the natural
scattering length case of Eq.~(\ref{Be}) and for the unnatural
case discussed in Sect.~\ref{unnat}.  Again, the only new issue is
the need to consider a particular spin state consistent with the
statistics of the problem.  If for example one is studying
identical spin 1/2 fermions---as in the case of two
protons---either the spin state is symmetric (spin one) and the
space is antisymmetric (odd L) or the spin state is antisymmetric
(spin zero) and the space symmetric (even L).  Since we have been
focusing on the case where $r_0 \gg R$ where only the s-wave
interaction is relevant the analysis of the previous sections
holds provided we restrict our attention to the total spin zero
sector.  This our previous analysis shows that two protons with
total spin zero will form a bound state in sufficiently strong
magnetic fields.

However, there is one critical way in which spin can become
important in this problem.  In free space it is always true that if
the two-particle system has a bound state, then the ground state
cannot be in the continuum. However, one cannot immediately
generalize the intuition gleaned from this fact to the case of bound
states in constant magnetic fields. Note that the characteristic
scale of the binding energy is of order $\omega$. The interaction
energy of the magnetic moments of particles with nonzero spin with
an external magnetic field are of the same characteristic order
(assuming the g-factors are of order unity). Moreover, for the case
of identical spin 1/2 particles these two effects are in
competition. The binding effect of the potential lowers the energy
for the spin zero state; the magnetic interaction lowers the energy
for the spin 1 state (with the moments aligned with the external
field). If this magnetic energy turns out to be larger then the
ground state is not the bound state.

As it happens, this is precisely the situation for the diproton. The
interaction energy of the magnetic moment of the two protons in a
spin  singlet state is, of course, zero.  In the spin aligned state
it is: \be
    \Delta E^{\rm spin-aligned} = -  g \, B_0 \,\mu_N = -\frac{ g \,\omega}{2} \ee
where $\mu_N$ the nuclear magneton is $\frac{e}{2 m}$.  The
g-factor for the proton is $g_p \approx 5.586$ yielding $\Delta
E^{\rm spin-aligned} \approx - 2.793 \, \omega$.  In contrast, one
sees from Fig.~\ref{fig_bindingenergy} that the binding energy is
bounded from above by $F_\infty \omega \approx .60486  \, \omega$.
Thus, the energy the system gained by binding in the spin singlet
channel is always more than four times smaller than the energy the
system gains by a spin flip into the spin aligned total spin 1
state.

The upshot of this is that although there is rather strong
evidence that the diproton has a bound state in sufficiently
strong field, the ground state will be in the two particle
continuum.

\section{Discussion}

As noted previously, the argument that in the regime where Coulomb
repulsion can be neglected attractive potentials for identical
charge particles will lead to bound states in magnetic fields is
quite general.  However, getting into the regime in which Coulomb
repulsion is negligible requires extraordinarily large magnetic
fields.  We see from Eq.~(\ref{deltaE_Cwc}) that unless there is
fine tuning of the scattering length, the Coulomb correction becomes
small when $\left ( Q^2 (\mu R) \frac{r_0^2}{R^2} \right ) \ll 1$. A
quick and dirty estimate for systems of atomic ions (with $R$ of
order the Bohr radius and $\mu \sim M_p$ gives $B_0 \gg 10^{11} \,
{\rm T}$. Clearly this is {\it far} beyond the regime for which the
ions are essentially unaltered from free space and the interaction
between the atoms is given by its free space value. The situation
only worsens for systems of molecular ions.

For this reason it is apparent that if the phenomenon occurs in a
clean system it is most likely to be for strongly interacting
particles. The diproton system is particularly favorable since the
scattering length is unnaturally large.  From Eq.~(\ref{deltaE_C}),
it is apparent that one requires magnetic fields of order $10^{12}
\, {\rm T}$ for the Coulomb repulsion to be modest enough to be
treated as a correction.   This scale is of physical interest.
Magnetic fields of this scale are believed to be found in
nature---in particular on the surface of magnetars.  Thus the novel
type of bound state discussed in this paper appears to be physically
relevant.

One important open issue associated with this phenomenon for the
diproton is whether there is a maximum magnetic field for which the
system binds.  As one increases the magnetic field there are
corrections to the effective field theory type treatment arising
from shorter distance effects; assuming such corrections are modest,
the scale of these corrections are given in Eq.~(\ref{deltaE_fr}).
As $B_0$ is increased further, however, these corrections will
become large and the expansion will break down. In principle one can
simply solve the problem using ``realistic'' nuclear potentials.
However, as $B_0$ increases, $r_0$ decreases. When $r_0$ becomes
short enough that it is similar to the spatial extent of the nucleon
one expects that magnetic effects would begin to cause substantial
modifications of the nucleon substructure; this in turn  could lead
to substantial modifications of the nucleon-nucleon potential. Since
the spatial extent of the nucleon is very similar to the range of
the nucleon-nucleon potential it is not clear whether an explicit
calculation with a realistic potential would provide any reliable
information beyond what can be studied via the effective theory
here.

Another important open question is whether there are any other
strongly interacting systems besides the diproton for which this
novel class of bound state will form.  As noted in Sect.~\ref{WF}
for systems with ``natural'' scattering lengths of order $R$ it
appears that typical nuclear systems are marginal at best.  If the
field is strong enough to overcome Coulomb repulsion, the system may
well be out of the weak field limit.  As the field increases beyond
this limit, the magnetic interactions may alter the internal
structure of the nuclei altering the potentials.  Thus, apart from
diprotons in magnetar strength fields we know of no other clean
examples of the phenomenon.

In looking for other systems, it should be noted that this paper
only considers the case for which the system consists of identical
charged particles; this leads to a decoupling of the center of mass
and relative motion. While this simplifies the problem both
conceptually and computationally, it also forces us to consider very
strong magnetic fields to overcome the effects of the Coulomb
repulsion. One might wish to consider the situation in which a
charged particle and a neutral particle interact in the presence of
a magnetic field. In such a situation, Coulomb repulsion would no
longer be an issue.  However, the notion of a two particle bound
state becomes a bit more involved in the presence of a magnetic
field since the relative and center-of-mass motion are coupled. In
practice, a useful definition is that the two particles are in an
energy eigenstate with the property that the position of the two
particles are strongly correlated---if one observes one of the
particles at some point, then the probability of not finding the
other particle a distance $r$ away decreases exponentially with $r$
for large $r$ in any direction.   An alternative definition is that
there exists a state with an energy less than the lowest Landau
level for the isolated charged particle.  The simple qualitative
arguments as to why one expects such bound states {\it a priori} are
weakened as compared to the case of identical charged particles; we
do not know whether magnetic fields can cause a charged and neutral
particle acting through a potential to bind (in the sense defined
above) when they do not bind in free space. We will consider this
possibility in a subsequent work.

Finally, we turn to the implications of this result from the
astrophysical perspective.  On the one hand, the magnetic field
strength in magnetars appears to be large enough for bound diprotons
to form.  An optimistic view on this would be that this can have
important effects on the dynamics of the magnetar leading to some
observational signature.  On the other hand, as discussed in
Sect.~\ref{spin}, the bound diprotons are not the ground state of
the system; spin aligned unbound protons are.  To the extent that
protons will tend to stay in their ground state, this suggests that
it will be difficult to find an observational signature of bound
diprotons.  Our results, however,  suggest that nuclei that are
unstable due to excessive number of protons may become stable in the
presence of a magnetic field of this magnitude. This shift of the
nuclear valley of stability towards higher values of $Z$ deserves
further study.


\end{document}